\documentclass[a4paper]{article} 

\def\comment#1{} 
\def\journalfont{\rm}         
\def\jou#1{{\journalfont #1\ }}
\def\joudef#1#2{\def #1{\jou{\ignorespaces #2}}}

\joudef{\aaa}    { Astron.\ Astrophys.}
\joudef{\aip}    { Adv.\ Phys.}
\joudef{\adm}    { adv.\ math.}
\joudef{\am}     { Ann.\ Math.}
\joudef{\apl}    { Ann.\ Phys.\ (Leipzig)}
\joudef{\apny}   { Ann.\ Phys.\ (N.Y.)}
\joudef{\apj}    { Astrophys.\ J.}
\joudef{\cjp}    { Can.\ J.\ Phys.}
\joudef{\cmp}    { Commun.\ Math.\ Phys.}
\joudef{\cqg}    { Class.\ Quantum Grav.}
\joudef{\grg}    { Gen.\ Rel.\ Grav.}
\joudef{\ijmpd}  { Int.\ J.\ Mod.\ Phys.\ D}
\joudef{\ijtp}   { Int.\ J.\ Theor.\ Phys.}
\joudef{\invm}   { Invent.\ Math.}
\joudef{\jm}     { J.\ Math.}
\joudef{\jmaa}   { J.\ Math.\ Anal.\ Appl.}
\joudef{\jmp}    { J.\ Math.\ Phys.}
\joudef{\jpa}    { J.\ Phys.\ A}
\joudef{\mnras}  { Mon.\ Not.\ R.\ Ast.\ Soc.}
\joudef{\mpla}   { Mod.\ Phys.\ Lett.\ A} 
\joudef{\nature} { Nature}
\joudef{\nc}     { Nuovo Cim.}
\joudef{\npb}    { Nuc.\ Phys.\ B}
\joudef{\ph}     { Physica}
\joudef{\pla}    { Phys.\ Lett. A}
\joudef{\plb}    { Phys.\ Lett. B}
\joudef{\pr}     { Phys.\ Rev.}
\joudef{\prd}    { Phys.\ Rev.\ D}
\joudef{\prep}   { Phys.\ Rep.}
\joudef{\prl}    { Phys.\ Rev.\ Lett.}
\joudef{\prsla}  { Proc.\ Roy.\ Soc.\ Lond.\ A}
\joudef{\ptp}    { Prog.\ Theor.\ Phys.}
\joudef{\ptps}   { Prog.\ Theor.\ Phys.\ Suppl.}
\joudef\rmp      { Rev.\ Mod.\ Phys.}
\joudef\spj      { Sov.\ Phys.\ JETP}

\catcode`@=11

\def\eqalign#1{\null\,\vcenter{\openup\jot\m@th
  \ialign{\strut\hfil$\displaystyle{##}$&$\displaystyle{{}##}$\hfil
      \crcr#1\crcr}}\,}
\def\meqalign#1{\null\,\vcenter{\openup\jot\m@th
  \ialign{\strut\hfil$\displaystyle{##}$&&$\displaystyle{{}##}$\hfil
      \crcr#1\crcr}}\,}
\catcode`@=12   

\newdimen\arrayruleHwidth
\makeatletter
\def\Hline{\noalign{\ifnum0=`}\fi\hrule \@height \arrayruleHwidth
  \futurelet \@tempa\@xhline}
\makeatother

\newcommand\thickbaselines{\baselineskip=20pt\lineskip=3pt\lineskiplimit=3pt}
\catcode`@=11
\def\cases#1{\left\{\,\vcenter{\thicknormalbaselines\m@th
             \ialign{$##\hfil$&\quad##\hfil\crcr#1\crcr}}\right.}
\def\matrix#1{\null\,\vcenter{\thickbaselines\m@th
    \ialign{\hfil$##$\hfil&&\quad\hfil$##$\hfil\crcr
      \mathstrut\crcr\noalign{\kern-\baselineskip}
      #1\crcr\mathstrut\crcr\noalign{\kern-\baselineskip}}}\,} 
\catcode`@=12   

\newcommand{\eprint}{\textsf} 

\newcommand\be{\begin{equation}} \newcommand\ee{\end{equation}} 
\newcommand\bd{\begin{displaymath}}\newcommand\ed{\end{displaymath}}

\newcommand\ts\textstyle

\def\undersim#1{\mathop{\vtop{\ialign{##\crcr
     $\hfil\displaystyle{#1}\hfil$\crcr\noalign
     {\kern1pt\nointerlineskip}\hbox{$\hfil\sim\hfil$}\crcr
     \noalign{\kern1pt}}}}}

\newcommand{\acronym}[3]{\newcommand{#1}{#3 (#2)\relax\renewcommand{#1}{#2}}}

 \def\etal{{\it et al.}}

\newtoks\reportnoregister \newtoks\eprintnoregister
\newcommand{\reportnumber}[1]{\reportnoregister={#1}}
\newcommand{\eprintnumber}[1]{\eprintnoregister={#1}}

\reportnumber{\mbox{}} 
\eprintnumber{\mbox{}} 

\newcommand{\reportid}{
   \begin{minipage}{17cm}\vspace{-3.2cm}
     \begin{flushright}
      {\normalsize \the\reportnoregister \\[-.2cm]
            \eprint{\the\eprintnoregister}}\vspace{3.2cm}
     \end{flushright}
   \end{minipage}\hspace{-17cm} }

\catcode`@=11   
\def\title#1{\gdef\@title{\reportid#1}}
\catcode`@=12   

\usepackage{amsmath}
\usepackage{graphicx}

\usepackage{amssymb}

\acronym{\SSS}{SSS}{{\em static spherically symmetric}}
\acronym{\TOV}{TOV}{{\em Tolman-Oppenheimer-Volkoff}}

\newcommand{\ncd}{\newcommand}
\ncd{\nms}{\negmedspace}
\ncd{\nts}{\negthickspace}
\ncd{\mcl}[1]{\mathcal{#1}}
\ncd{\beq} {\begin{equation}}
\ncd{\eeq} {\end{equation}}
\ncd{\BE} {\begin{eqnarray}}
\ncd{\EE} {\end{eqnarray}}
\ncd{\rarr} {\rightarrow}
\ncd{\larr} {\leftarrow}
\ncd{\lrarr} {\leftrightarrow}
\ncd{\lbeq}[1]  {\label{eq: #1}}
\ncd{\refeq}[1] {(\ref{eq: #1})}
\ncd{\mrm}    {\mathrm}
\ncd{\nn}{\nonumber}
\ncd{\mbf}[1] {{\mathbf #1}}
\ncd\T{\frac{1}{2}h^{\mu\nu}p_\mu p_\nu}
\ncd{\ms}{\mathstyle}
\ncd{\ds}{\displaystyle}
\ncd{\Yc}{Y_{\rm c}}
\ncd{\Yorb}{Y_{\rm orb}}

\ncd{\der}{{\mathrm{d}}}
\ncd{\rtil}{\tilde{r}}
\ncd{\Mr}{\frac{2M}{r}}
\ncd{\rhotil}{\tilde{\rho}}

\begin{document}

\reportnumber{USITP 2000-14}

\title{\Large Constructing stellar objects with multiple necks}
  \author{Max Karlovini\footnote{E-mail: \eprint{max@physto.se}}\;, 
     Kjell Rosquist\footnote{E-mail: \eprint{kr@physto.se}}\; and 
     Lars Samuelsson\footnote{E-mail: \eprint{larsam@physto.se}} \\[10pt]
  {\small Department of Physics, Stockholm University}  \\[-10pt]
  {\small Box 6730, 113 85 Stockholm, Sweden} \\
\begin{minipage}[t]{0.8\linewidth}\small{ We discuss the construction 
    of perfect fluid stellar objects having
    optical geometries with multiple necks corresponding to spatially
    closed unstable lightlike geodesics. We prove that there exist
    physically reasonable models with arbitrarily many necks. We also
    show how a first order phase transition can give rise to quite
    pronounced secondary double
    necks. The analysis is carried out using a modification of a
    recent dynamical systems formulation of the TOV equations due to
    Nilsson and Uggla. Our reformulation allows for a very general family of
    equations of state including, for example,  phase transitions. 
 }\end{minipage}}

\date{}

\maketitle


\section{Introduction}
The theory of \SSS\ relativistic stellar models has received an increasing
interest in recent years.  A few years ago Chandrasekhar and Ferrari
\cite{cf:gwresonance} discovered the phenomenon of trapped gravitational waves
in ultracompact objects.  Later Abramowicz \etal\ \cite{aabgs:optical} showed
that the trapping could be conveniently visualized in terms of the optical
geometry of the models.  It follows from their work that trapping occurs when
the optical geometry has a sufficiently pronounced neck.  The physical
relevance of the neck stems from the fact that it is associated with an
unstable closed lightlike circular orbit.  This phenomenon is present in the
exterior Schwarzschild geometry at $R=3M$ as is well-known.  Any \SSS\ stellar
object which is more compact than $R=3M$ will therefore have a neck.  However,
Rosquist showed in \cite{rosquist:trapped} that the limit $R\lesssim 3M$ is not
essential for the formation of a neck and hence it is not essential for a
trapping region either.  In particular, a neck can exist in the stellar
interior.  Because the optical geometry is regular all the way down to the
center of the star, a neck is always associated with a bulge which is located
further inside the stellar interior.  Moreover, as was pointed out in
\cite{rosquist:multneck}, there exist models with multiple
necks.  In fact the number of necks can be unbounded for a given equation of
state.  Some examples of double neck models with quite pronounced
double necks were given in \cite{krs:annalen}. In the present paper we
show how such models can be constructed.  We also prove the statements
made in \cite{rosquist:multneck}.

The relation between necks and the optical geometry can be described in terms
of the centrifugal potential ($V_l$) which governs the null geodesic orbits in
the stellar spacetime.  A slight modification ($V=V_l+V_m$) of this potential is
responsible for the propagation of axial gravitational wave modes.  In this
picture a bulge corresponds to a potential well in which trapped modes may
exist.  Models with multiple necks (and hence bulges) have potentials with
multiple wells.  The corresponding gravitational wave modes can therefore be
considerably more complicated than the single well modes.

Our analysis relies on a recent dynamical systems approach by Nilsson and
Uggla \cite{nu:grstars_linear, nu:grstars_polytropic} who treated the full
solution space of the \SSS\ field equations, with focus on the case of linear
and polytropic equation of state.  Nilsson and Uggla give three different
dynamical systems reformulations of the field equations, one for the linear
case \cite{nu:grstars_linear} and two complementary ones for the polytropic
case \cite{nu:grstars_polytropic}.  Referring to the shape of the
corresponding state spaces, we denote these formulations as the saddle
system\footnote{The ``saddle'' here refers to the shape of the surface of
vanishing pressure (which serves as a boundary for physical solutions), rather
than the shape of the full state space.} \cite{nu:grstars_linear}, the loaf of
bread system \cite{nu:grstars_polytropic} and the cube system
\cite{nu:grstars_polytropic}.

Here, we will be using a variant of the loaf of bread formulation to prove the
existence of optical geometries having an arbitrarily large number of multiple
necks \cite{rosquist:multneck}.  Surprisingly this phenomenon occurs for the
extremely simple Zel'dovich stiff matter equation of state with an added
constant, $p=\rho -\rho_\mathrm{s}$ (we use the ``s'' subscript to denote
evaluation at the stellar surface where $p=0$).  This equation of state is
everywhere causal, the speed of sound being exactly equal to the speed of
light.  However, the necks are quite unobtrusive and apparently do not give
rise to trapped gravitational wave modes.  Therefore, for the sake of
illustration, we use models with a first order phase transition in this paper.
For simplicity we focus on models with two uniform density layers.  Such
models were recently considered by Lindblom \cite{lindblom:phase} who 
studied their mass-radius
curves.  It turns out that the phase transition has a quite dramatic effect on
the gravitational field in the stellar interior.  In particular it is possible
to achieve very pronounced necks if the phase transition is sufficiently
strong \cite{krs:annalen}.

We are leaving issues of stability aside in this paper.  However, we may
remark that the Zel'dovich equation of state is in fact unstable beyond the
first neck as can be seen from its mass-radius curve.  The stability of the
two-zone uniform density models remains unclear.  The mass-radius curves
cannot be used since there is no well-defined adiabatic index given by the
equation of state itself.  Therefore a dynamical stability analysis must be
employed to determine the stability.  However, there is a technical difficulty
here since the perturbation equation is not of the Sturm-Liouville type
because of the discontinuity in the density function.  Hence the standard
theorem for the spectrum of eigenfrequencies does not apply in this case.

\section{The modified loaf of bread system with the inverse chemical potential as matter variable} 
\label{sec:regularized}
In the Schwarzschild radial gauge, the metric for a static spherically
symmetric (SSS) spacetime can be written as
\begin{equation} \label{SSSmetric}
  ds^2 = -e^{2\nu}dt^2 + e^{2\lambda}dr^2 +
  r^2(d\theta^2+\sin^2\theta\,d\phi^2)
\end{equation} 
with $\nu$ and $\lambda$ being functions of the radial variable only.
In the case of an isentropic perfect fluid, described by a pressure
$p$ and an energy-density $\rho$ constrained by an equation of state
$\rho=f(p)$, the stress-energy tensor takes the
form
\begin{equation} 
  T_{ab} = \rho\,u_a u_b + p\,(g_{ab} + u_a u_b),
\end{equation}
where $u^a$ is the four-velocity of the fluid. The SSS symmetry
implies that $u^a = e^{-\nu}(\partial/\partial t)^a$ and that $p$ and
$\rho$ are functions of the radius only. We shall restrict our
attention to fluids satisfying the basic physical requirements that
$p$, $\rho$ and $d\rho/dp$ are everywhere non-negative. We will also
assume that the equation of state has the following asymptotic
behaviours
\begin{equation}
\begin{align}\lbeq{pzero}
p/\rho  &\rarr 0 \quad \mbox{ as } p\rarr 0 \\ \lbeq{pinfty}
\rho/p  &\rarr \mrm{constant} \geq 0 \quad \mbox{ as } p\rarr
\infty 
\end{align}
\end{equation}
The $p\rarr 0$ behaviour imposes no restriction on stellar models,
since it is necessary \cite{lm:indexlimit} for the existence of
solutions with finite radii.  The $p\rarr\infty$ behaviour could also
be expressed as saying that the speed of sound $v_{\rm sound} =
\sqrt{dp/d\rho}$ should not go to zero (or oscillate) in the given
limit. Furthermore, if $p/\rho$ were to vanish in the large pressure
limit (so that $\rho/p$ diverges), solutions with large central
pressures would be Newtonian ($p\ll\rho$) near the center, while, on
the contrary, a large central pressure star is expected to be highly
relativistic near its center. 

For future reference we note that condition \refeq{pinfty} implies
that the equation of state can be uniquely split in the useful form  
\begin{equation}
\rho = (\eta-1)p + \psi,
\end{equation}
where $\eta=1+\lim_{p\rarr\infty}\rho/p \geq 1$ and $\psi$ is a
function of $p$ satisfying $\psi/p \rarr 0$ as $p\rarr\infty$.
Furthermore eq.\ \refeq{pzero} implies that $p/\psi\rarr 0$ as $p\rarr
0$. There are further physical conditions that can be imposed,
specifically the dominant energy condition $p\leq \rho$, as well as
the condition that the fluid should not become superluminal, which is
saying that the velocity of sound $v_{\mrm{sound}}$
should not exceed the value $1$.  However, these two conditions need
not be stressed in this general discussion, but note that they
independently imply $\eta\geq 2$.

A global picture of the solution space of Einstein's equations
$G_{ab}=\kappa T_{ab}$ for such a model can be obtained by formulating
the equations as a dynamical system, preferably on a compact state
space using dimensionless variables. In this work we use the loaf of
bread formulation of Nilsson and Uggla as our starting point. Later on
we will modify that formulation to better suit our purposes. In the
loaf of bread formulation Nilsson and Uggla use two geometric
variables given by
\begin{equation}
\begin{align}\lbeq{Sigmadef}
  \Sigma &= \frac{r\nu_{,r}}{1+r\nu_{,r}} \\ \lbeq{Kdef} 
       K &= \frac{e^{2\lambda}}{(1+r\nu_{,r})^2}\,,
\end{align}
\end{equation}
where $\nu_{,r}=d\nu/dr$. In these variables, the quotient $m/r$ is given
by 
\begin{equation}\lbeq{mr}
  1-\frac{2m}{r} = e^{-2\lambda} = \frac{(1-\Sigma)^2}{K},
\end{equation}
where $m$ is the standard spherically symmetric mass function of
Misner and Sharp \cite{ms:msmass}, which can be invariantly defined in
terms of the Schwarzschild radius $r$ by the relation
\begin{equation}
  1-\frac{2m}{r} = \nabla^a r\,\nabla_{\!a}r.
\end{equation}
For the matter degree of freedom, Nilsson and Uggla chose
\begin{equation}
y = \frac{p}{\rho+p}\,.
\end{equation}
The independent variable $x$ is defined up to translations by
\begin{equation}\lbeq{xdef}
dx = \frac{dr}{y(1-\Sigma)r}\,.
\end{equation}
This gives the evolution equations
\begin{equation}
\begin{align}\lbeq{Sigmadot}
  \Sigma' &= -yK\Sigma +\ts\frac12 P[1+2y(1-2\Sigma)] \\ 
  \lbeq{Kdot} K' &= 2y(1-K-2P)K \\ 
  \lbeq{ydot} y' &= -yF(y)\Sigma
\end{align}
\end{equation}
where $P := 1-\Sigma^2-K$, $F(y) := y(\rho/p-d\rho/dp) =
y(\psi/p-d\psi/dp)$ and a prime denotes differentiation with respect to
$x$. This formulation is based on the assumption that the function $F$
is strictly positive except at infinite pressure ($y=\eta^{-1}$) where
it becomes zero. When the assumption holds, the boundaries of the
state space are defined by the following invariant submanifolds
\begin{equation}
\begin{align} 
  y &= 0  \quad        \mbox{(surface of zero pressure)} \lbeq{sub1} \\
  y &= \eta^{-1} \quad \mbox{(surface of infinite pressure)}  \\ 
  P &= 0 \quad         \mbox{(i.e.\ $K = 1-\Sigma^2$)} \lbeq{sub3} \\
  K &= 0 \,.  \lbeq{sub4}
\end{align}
\end{equation}
This clearly gives a compact state space, whose coordinate image in
$\mathbb{R}^3$ is shaped like a loaf of bread (see fig.\ 
\ref{fig:stiff_loaf}). There are no equilibrium points in the interior
of the state space, implying that the topological orbit structure is
completely determined by the equilibrium points on the boundary
surfaces. The equilibrium points and their associated eigenvalues are
given in \cite{nu:grstars_polytropic} for the case of a polytropic
equation of state. The evolution equation for the pressure being
$\dot{p}=-p\Sigma$, the right half of the state space defined by
$\Sigma\geq 0$ is the physical one where a positive pressure is
non-decreasing. A physical solution is furthermore required to be
regular at the center of the star, which implies that the center
corresponds to a point on the ``ridge'' ($\Sigma = 0$, $K = 1$) of the
loaf. The surface of the star corresponds to a point on the arc $y=0$,
$P=0$, $\Sigma \geq 0$, which serves as an attractor for all interior
points of the state space. By the definition of the
independent variable $x$, an orbit representing a physical star
approaches a point on the ridge as $x\rarr -\infty$ and a point on the
surface arc as $x\rarr \infty$. By evaluating eq.\ \refeq{mr} on the
surface arc where $K=1-\Sigma^2$, one finds $M/R = m_{\mathrm
  s}/r_{\mathrm s}$ being given by
\begin{equation}\lbeq{MR}
  \frac{M}{R} = \frac{\Sigma_{\mathrm s}}{1+\Sigma_{\mathrm s}},
\end{equation}
which determines the matching to the exterior Schwarzschild geometry.
An orbit for which $\Sigma_{\mathrm s} = 0$ has an infinite radius
$R$, but could still have a finite mass $M$, but here we do not address
this issue in detail since we shall only be interested in models with
finite radii.

According to eq.\ \refeq{ydot}, the assumption of $F$ being positive
for finite pressures means that $y$ is a monotone variable. However,
although eqs.\ \refeq{pzero} and \refeq{pinfty} ensure that $F$ will
be positive in the limit of small as well as large (but finite)
pressures, a physically reasonable equation of state can have a soft
intermediate regime where $F$ drops below zero, a notable example
being the Harrison-Wheeler equation of state \cite{htww:gravcollapse}.
For such equations of state, this formulation is only suited to
describe stars for which the central pressure $p_{\rm c}$ is less than the
lowest pressure $p_1$ which gives $F=0$. The reason is that every
pressure $p_k$ for which $F=0$ defines an invariant submanifold
$y=y_k$. Hence for a star with $p_{\rm c}\geq p_1$, the independent variable
$x$ will run off to infinity before the surface arc is reached. 

To remedy this situation we shall use a different matter variable,
namely one with a simple relation to the manifestly monotone, bounded
and dimensionless function
\begin{equation}\lbeq{hdef}
  w = e^{-h},
\end{equation}
where
\begin{equation}
h = \int_0^p\frac{dq}{f(q)+q}.
\end{equation}
Note that $h$ has a simple relation to the gravitational
potential, $\nu=-h+constant$. By its definition $h$ is a continuous
function of the pressure whether or not
the same holds true for the density $\rho=f(p)$. Hence, by using a
continuous function of $h$ as matter variable, we will also be able to
treat equations of state with a phase transition $\lim_{p\uparrow
  p_{\rm t}}\rho < \lim_{p\downarrow p_{\rm t}}\rho$ at one or several pressure
values $p_{\rm t}$. 

The function $w$ has a simple physical interpretation of being
proportional to the inverse of the chemical potential, which often is
defined as \cite{mtw:gravitation} 
\begin{equation}
  \mu = \frac{\rho+p}{n_{\rm p}},
\end{equation}
where $n_{\rm p}$ is the particle number density. It then follows from
the relation $n_{\rm p}=(d\mu/dn_{\rm p})^{-1}$ \cite{mtw:gravitation}
that the chemical potential may be reexpressed as $\mu = \mu_{\mrm s}
w^{-1}$. 

For the equations of state we consider, the function $h$ has the
asymptotic behaviours
\begin{equation}
\begin{align}\lbeq{hzero}
  h\rho/p &\rarr 1 \quad \mbox{as $p\rarr 0$} \\ \lbeq{hinfty}
  h/\ln{p} &\rarr \eta^{-1} \quad \mbox{as $p\rarr\infty$} 
\end{align}
\end{equation}
Since by assumption $p/\rho\rarr 0$ as $p\rarr 0$, eq.\ \refeq{hzero}
shows that although the integrand of eq.\ \refeq{hdef} may be
divergent at $p=0$, the integral itself is well-behaved. It follows
that $w$ has the compact range $0\leq w\leq 1$. There are, however,
two reasons why we shall not choose to use $w$ itself as the matter
variable for the dynamical system. Firstly, we want the matter
variable to decrease with decreasing pressure and to become zero at
$p=0$. This will make our reformulation qualitatively more similar to
the original formulation which uses $y=p/(\rho+p)$. The requirement can
of course always be trivially met, e.g.\ by using $1-w$ rather than
$w$. Secondly, if the matter variable is not chosen properly, the
dynamical system will in general become irregular on the surface of
infinite pressure. Since there will be equilibrium points on this
surface, important for the solution structure, this would be an
unwanted feature.

Referring to our sought for matter variable as $Y$, its evolution
equation can be written as 
\begin{equation}\lbeq{Ydotprel}
  Y' = yw\,\frac{dY}{dw}\Sigma.
\end{equation}
Combined with the evolution equations \refeq{Sigmadot} and
\refeq{Kdot} for the geometric variables, we see that the system will
be linearizable if $y$ and $w\,dY/dw$ are $C^1$ functions of
$Y$. According to eq.\ \refeq{ydot} and \refeq{Ydotprel}, we have
\begin{equation}
  \frac{dy}{dY} = -\frac{F}{w\,dY/dw} =
  -y\frac{\Delta}{w\,dY/dw},
\end{equation}
where $\Delta=\psi/p-d\psi/dp$. Assuming that $w\,dY/dw$ is nonzero at
finite pressures, the above expression for $dy/dY$ stays finite
for finite pressures, but is clearly at risk 
of becoming divergent at infinite pressure, where $y=\eta^{-1}$ and
$\Delta=w=0$. Hence we must choose $Y$ such that the quotient
\begin{equation}
Q=\frac{\Delta}{w\,dY/dw}
\end{equation}
stays finite as $p\rarr\infty$. In order to avoid equilibrium points
with zero eigenvalues, $Q$ should preferably take a nonzero value in
this limit. For simplicity we shall only consider the case when the
leading behaviour of $\psi$ at high pressures is constant or
polytropic, i.e. we shall assume that there exists a nonnegative
finite number $n$ such that
\begin{equation}\lbeq{defn}
  \psi p^{-n/(n+1)} \rarr \mrm{constant}\neq 0 \quad \mbox{ as }
  p\rarr\infty. 
\end{equation}
With this assumption, $\Delta$ and $w$ will satisfy
\begin{equation}
\begin{align}
  &\Delta p^{\,1/(n+1)} \rarr \mrm{constant}\neq 0 \\
  &w p^{\,1/\eta} \rarr \mrm{constant}\neq 0,
\end{align}
\end{equation}
in the same limit. It follows that we should not use a linear function
of $w$ as the matter variable unless $n+1=\eta$. However, the choice
\begin{equation}\lbeq{Ydef}
  Y = 1 - w^s, \quad s = \frac{\eta}{n+1},
\end{equation}
gives $w\,dY/dw = -sw^s = -s(1-Y)$ and hence a quotient $Q$ which
stays finite and nonzero as $p\rarr\infty$. 

Finally settling for this choice, the evolution equations for the
dynamical system take the form
\begin{equation}
\begin{align}\lbeq{Sigmadot2}
 \Sigma' &= -yK\Sigma+\ts\frac12 P[1+2y(1-2\Sigma)] \\ \lbeq{Kdot2}
 K' &= 2y(1-K-2P)K \\ \lbeq{Ydot}
 Y' &= -\frac{\eta}{n+1}\,y(1-Y)\Sigma,
\end{align}
\end{equation}
where $y$ should now be viewed as a function of $Y$. Strictly speaking
this system is not a proper dynamical system if there is a phase
transition. In that case $y$ has a discontinuity which leads to
discontinuities in the derivatives of all the state space
variables. This is however not a serious problem as long as the
numerics is done carefully near the phase transition. 

It may be noted that for an
equation of state of the type $\rho = (\eta-1)p + kp^{\,n/(n+1)}$,
i.e.\ $\psi = kp^{\,n/(n+1)}$, our matter variable $Y$ can be explicitly
expressed as a function of $p$ as
\begin{equation}
  Y = \frac{v}{1+v}, \quad v = \frac{\eta}{k}\,p^{\,1/(n+1)},
\end{equation}
In fact, this shows that $Y$ is then proportional to the original
matter variable $y$ according to $Y=\eta y$. This particular class of
equations of state includes the incompressible case ($\eta=1$, $n=0$),
the polytropic case ($\eta=1$, $n>0$), the relativistic polytrope
($\eta=n+1$, $n>0$) as well as the linear case ($n=0$, $k=\rho_{\rm
  s}$).
More generally, for every equation of state that satisfies our general
assumptions \refeq{pzero}, \refeq{pinfty} and \refeq{defn}, the qualitative
picture of the state space resembles that
of the original formulation for the case of $y$ being monotone.
Quantitatively, the invariant boundary submanifolds $y=0$ and
$y=\eta^{-1}$ are replaced by $Y=0$ and $Y=1$, respectively. We stress
that there will be no additional submanifolds even if $y$ is
nonmonotone.  Moreover, the reformulation alters the eigenvalues and
eigenvector components of the equilibrium points. The new eigenvalues
of all possible equilibrium points are given in table
\ref{table:Yeig}.

As far as the numerical integration of the system given by eq.\ 
\refeq{Sigmadot2}, \refeq{Kdot2} and \refeq{Ydot} is concerned, one
encounters a difficulty when using an equation of state for which it
is not possible to explicitly express $y$ as a function of $Y$, since
$y$ has explicit occurrence in the left hand sides of the equations. In
the case when the equation of state is given in parametrized form,
\begin{equation}
  p=p(\lambda), \quad \rho=\rho(\lambda),
\end{equation}
with $p$ and $\rho$ being monotone functions of the
parameter $\lambda$, one can easily get around the difficulty by
viewing $y$ as a function of $\lambda$ and utilizing $\lambda$ as an
auxiliary dynamical variable with the evolution equation
\begin{equation}
  \lambda' = -p\left(\frac{dp}{d\lambda}\right)^{\!\!\!-1}\!\Sigma\,.
\end{equation}
This gives two possibilities of calculating $Y$. Either one can
integrate the three evolution equations for $\Sigma$, $K$ and
$\lambda$ and, given the result, calculate $Y$ according to
\begin{equation}\lbeq{Ylambda}
   Y = 1 - \exp \left\{-\frac{\eta}{n+1}\int_{\lambda_{\mrm{s}}}^\lambda
   \frac{dp/d\lambda}{\rho+p}d\lambda\right\}, 
\end{equation}
or else one can integrate the evolution equations for $\Sigma$, $K$,
$Y$ and $\lambda$ simultaneously. In the latter case eq.\ 
\refeq{Ylambda} plays the role of a propagated constraint, which
offers the possibility of putting the numerics to a test by afterwards
checking the accuracy to which the relation holds.

For each equation of state, there is a one-parameter family of stars
corresponding to the choice of central pressure $p_{\mrm c}$. This
choice is the sole initial condition, specifying the starting point
$Y=Y_{\mrm c}$ up on the ridge of the state space. However, since all
points on the ridge are equilibrium points, the starting point of the
actual integration has to be moved out a tiny distance from the ridge
point in the eigendirection of the positive eigenvalue $2y_{\mrm c}$.
This point is given by the following linearized solution near the
center:
\begin{equation}
\begin{align}\lbeq{Sigmalin}
  \Sigma &= 2(1+2y_{\mrm c})\epsilon\,e^{2y_{\mrm c}x} \\ \lbeq{Klin}
  K &= 1 - 12y_{\mrm c}\epsilon\,e^{2y_{\mrm c}x} \\ \lbeq{Ylin}
  Y &=  Y_{\mrm c}-s(1-Y_{\mrm c})(1+2y_{\mrm c})\epsilon\,e^{2y_{\mrm c}x},
\end{align}
\end{equation}
where $\epsilon$ is the trivial integration constant corresponding to the
freedom to make translations in $x$. The corresponding linearization
for the auxiliary variable $\lambda$ is 
\begin{equation}
  \lambda = \lambda_{\mrm c} - \frac{1+2y_{\mrm c}}{y_{\mrm c}}p_{\mrm
  c}\!\left(\frac{dp}{d\lambda}\right)_{\!\!\mrm c}^{\!\!-1}\epsilon\,e^{2y_{\mrm c}x}.
\end{equation}
\section{The surface of regular orbits}
\label{sec:regular}
The orbits with regular centers, i.e.\ the orbits starting out on the
ridge $\Sigma = 0$, $K = 1$ and leaving it according to the linearized
solution given by eqs.\ \refeq{Sigmalin} - \refeq{Ylin}, defines a
subset $V$ of the state space $S$ which we will here discuss in some
detail. The fact that the interior of $S$ contains no equilibrium
points implies that all orbits begin and end at
equilibrium points on the boundary $\partial S$. As a consequence, the
interior of $V$ is an embedded two-dimensional submanifold in the
three-dimensional interior of $S$. To find out how this two-surface is
embedded, it is clearly important to understand the way the boundary
$\partial V$ is situated in $S$. To this end we first need to
understand the stability properties of the subset of equilibrium
points in $\partial S$ which also belongs to $\partial V$. This set of
critical points is comprised of the line of stellar center points
$L_2$, the single point $P_1$ on the surface of infinite pressure
$Y=1$ as well as the line of stellar surface points $\tilde{L}_1$ (the
part of the line of equilibrium points $L_1$ for which
$\Sigma\leq\Sigma_{\rm max}$ for some $\Sigma_{\rm max}\leq 4/5$ which
according to eq.\ \refeq{MR} gives the minimal value of the tenuity
$R/M\geq 9/4$). The eigenvalues of these points as well as the exact
position of $P_1$ can be found in table \ref{table:Yeig} and are shown 
in figure \ref{fig:stiff_loaf}.

Let us first discuss the line of points $L_2$, which is the part of
$\partial V$ from which the regular orbits start out. Setting $Y=0$
gives the point which connects $L_2$ with $\tilde{L}_1$ and simply
corresponds to the Minkowski solution. As can
be read off from table \ref{table:Yeig}, this point is non-hyperbolic,
i.e.\ all three eigenvalues vanish implying that the orbital structure
near this point can not be obtained by a linearization of the system
of ODEs.  For $0<Y<1$, points on $L_2$ has one zero, one positive and
one negative real eigenvalues. While the negative eigenvalue is
associated with an eigenvector tangential to the boundary surface
$P=0$, the eigenvector of the positive eigenvalue, which generates the
regular orbits and can be found in table \ref{table:eigvec}, is
directed into the $\Sigma>0$ half of the interior of $S$ and towards
decreasing values of $Y$. However, when going to infinite central
pressure by setting $Y=1$, the $Y$ component of the latter eigenvector
vanishes and generates an orbit which necessarily is lying in the
invariant submanifold $Y=1$ and is ending at $P_1$. Since this orbit
never reaches $L_1$, it obviously corresponds to a stellar model which
is infinite in extent. Indeed, since the form of the pressure function
$\psi$ in the equation of state $\rho=(\eta-1)p+\psi$ is irrelevant as
long as one stays on $Y=1$, it gives a solution to the Einstein
equations which can be identified with any regular center solution
that is obtained for the scale invariant equation of state
$\rho=(\eta-1)p$. This follows from the fact that all regular center
solutions for the scale invariant equation of state are equivalent as
they simply are related by a rescaling of dimensionful variables, a
change of units if one prefers, and hence they are all equivalent to
the infinite central pressure limit solution. For this reason, we
shall refer to the orbit in question as the RSI (regular scale
invariant) orbit. Note that the RSI orbit is a part of $\partial V$,
connecting $L_2$ with $P_1$.

As far as the equilibrium $P_1$ is concerned, this point itself again
corresponds to a solution for the scale invariant equation of state,
namely the self-similar Tolman solution, which neither has a regular
center nor a finite radius. The point $P_1$ has two eigenvalues whose
real parts are always negative and whose eigenvectors are tangential
to the $Y=1$ surface. For $\eta<\eta_*$, $\eta=\eta_*$ and
$\eta>\eta_*$ with $\eta_* = -2+8/\sqrt{7}\approx 1.024$, these
eigenvalues are real and distinct, real and equal respectively complex
conjugate. In particular, this means that for $\eta>\eta_*$ the RSI
orbit will spiral around $P_1$ infinitely many times before reaching
it. The third eigenvalue is always real and positive and the
associated eigenvector, which is given in table \ref{table:eigvec},
points into the interior of $S$. It generates an orbit corresponding
to a solution which does not have a regular center but which does end
on the stellar surface line $\tilde{L}_1$.  This orbit will be refered
to as the skeleton orbit. Just like the RSI orbit, the skeleton orbit
is part of $\partial V$, this time giving the connection between $P_1$
and $\tilde{L}_1$. It is important to understand that one may view the
\emph{combined} RSI and skeleton orbits, joined together at $P_1$, as a
limiting orbit for the regular center orbits: by choosing a finite but
sufficiently large central pressure, it is possible to obtain a
regular orbit which follows this combined limiting orbit arbitrarily
closely. In particular, let us assume that $\eta>\eta_*$ and focus on
the projection of finite central pressure orbits onto the $Y=1$ plane.
Since the RSI part of the combined limiting orbit in this case is a
spiral, there can be no upper limit on the number of times such an
orbit may spiral around the point $P_1$, given that $\eta>\eta_*$. As
we shall see, this is closely related to the possibility of
constructing stellar models with an arbitrarily large number of necks
in the optical geometry.

We now finally turn to $\tilde{L}_1$. Equilibrium points on $L_1$ with
$\Sigma>0$ have one zero and two negative eigenvalues, $\lambda_1 =
-\Sigma$ and $\lambda_2 = -F_{\rm s}\Sigma$, which are distinct unless
the matter function $F = y(\psi/p-d\psi/dp)$ takes the value $1$ at
zero pressure. Reexpressing the surface value of $F$ as $F_{\rm s} =
1 - \lim_{p\rarr 0}(p/\rho)(d\rho/dp)$ reveals that $0\leq F_{\rm
  s}\leq 1$ and that $F_{\rm s}$ only depends on the leading behaviour
of $\psi$ in the limit of small pressures. We note that if the leading
behaviour is polytropic, i.e.\ if there is some number $\tilde{n}\geq
0$ such that $\psi p^{-\tilde{n}/(\tilde{n}+1)}\rarr\rm const >0$ as
$p\rarr 0$, $F$ will evaluate to the number $1/(\tilde{n}+1)$ at zero
pressure and hence will take the value $1$ if $\tilde{n}=0$. We note
in passing that the polytropic type index $\tilde{n}$ in general need
not coincide with the index $n$ defined by eq.\ \refeq{defn}.
Moreover, we remark that it is not true that $F_{\rm s} = 1$ iff the
surface density $\rho_{\rm s}>0$, since e.g.\ a leading behaviour of
the perhaps unrealistic type $\psi\ln{\!(p_1/p)}\rarr\rm const > 0$
as $p\rarr 0$ gives $F_{\rm s} = 1$. If $F_{\rm s}<1$ so that
$\lambda_1>\lambda_2$, the eigenvector $V_1$ belonging to $\lambda_1$
is directed in the pure $\Sigma$ direction, while the eigenvector
$V_2$ of $\lambda_2$ is instead tangential to the state space boundary
$P=0$, its $K$ component being $-2\Sigma$ times its $\Sigma$
component. From table \ref{table:eigvec} it can be read off that the
$\Sigma$ component has the same sign as the $Y$ component, which
should be chosen negative for the eigenvector to point towards
decreasing pressure. However, since all eigenvectors are here found to
be tangential to state space boundaries which are invariant
submanifolds, no interior orbit can enter a point on $\tilde{L}_1$
\emph{exactly} along an eigenvector direction. Setting $F_{\rm s} = 1$
so that $\lambda_1 = \lambda_2$, the two eigenvectors $V_1$ and $V_2$
are obviously still linearly independent which implies that every
linear combination of these two vectors is also an eigenvector. In
this case it is impossible to use the eigenvector structure to draw
any conclusions about how the regular orbits enters $\tilde{L}_1$,
except for the fact that they must enter tangentially to the plane
spanned by $V_1$ and $V_2$. Some numerical studies using the equation
of state $\rho=(\eta-1)p+kp^{\,n/(n+1)}$ for which $F_{\rm s}=1/(n+1)$
seem to indicate the following. For $F_{\rm s}$ equal or close to
unity, the regular orbits in general do not tend to enter
$\tilde{L}_1$ along neither $V_1$ nor $V_2$. However, with increasing
$F_{\rm s}$, the orbits tend to approach $\tilde{L}_1$ tangentially to
the $P=0$ surface, along the eigenvector direction provided by $V_2$.
This is to be expected, since the eigenvalue quotient
$\lambda_2/\lambda_1 = F_{\rm s}$ decreases with $F_{\rm s}$.
Moreover, with increasing $n$, the orbits will tend to end at smaller
values of $\Sigma$ until eventually all orbits will end at the
non-hyperbolic Minkowski point at $\Sigma=0$ where the local analysis
breaks down.

Now, an important point regarding the $\tilde{L}_1$ part of $L_1$, is
that its points need not be in one to one correspondence with the
regular center orbits. Since the value of $\Sigma$ on $\tilde{L}_1$ by
eq.\ \refeq{MR} determines the compactness $M/R$, this is directly
related to the fact that the mass-radius curve associated with a given
equation of state does not in general define a monotone relation
between $M$ and $R$. In particular, the maximal surface value
$\Sigma_{\rm max}$ (and hence the maximal compactness $M/R$) need not
be the one obtained for the skeleton orbit, i.e.\ when going in the
limit of infinite central pressure. Indeed, for many equations of
state the mass-radius curve ends in a spiral, corresponding to the
surface of regular orbits being rolled around its skeleton orbit
boundary, always without crossing itself, before it is finally
squashed into the $\tilde{L}_1$ line. As we have seen, when
$\eta>\eta_*$, the surface is already rolled around the skeleton orbit
when leaving the $Y=1$ surface, so it seems likely that the high
pressure regime of the mass-radius curve is closely related to the
structure of the RSI orbit.


\begin{table}[ht]
\begin{tabular}{|c|ccc|c|c|} \hline\hline
Eq.\ point &  $\Sigma$  &  $K$  &  $Y$  &  Eigenvalues \\
\hline
$L_1$  & $\Sigma$  &  $1-\Sigma^2$  &  $0$  & $-\Sigma$, $0$, $-F\Sigma$ \\[-6pt]

$L_2$  &  $0$  &  $1$  &  $Y$  &  $0$, $-y$, $2y$ \\[2pt]

$P_1$  &  $\ds\frac{2}{\eta+2}$  &  $\ds\frac{\eta^2+4\eta-4}{(\eta+2)^2}$  &
$1$  &  $\ds\frac2{(n+1)(\eta+2)}$, $\ds\frac1{2\eta}\left(-1 \pm
  i\frac{\sqrt{7\eta^2+28\eta-36}}{\eta+2}\right)$ \\[7pt]

$P_2$  &  $1$  &  $0$  &  $1$  &  $\ds\frac2{\eta}$, $\ds\frac1{n+1}$, $-\ds\frac{\eta-2}{\eta}$ \\[4pt]

$P_3$  &  $-1$  &  $0$  &  $1$  &  $\ds\frac2{\eta}$, $-\ds\frac1{n+1}$, $\ds\frac{\eta+6}{\eta}$ \\[5pt]

$P_4$  &  $\ds\frac{\eta+2}4$  &  $0$  &  $1$  &
$\ds\frac{\eta^2+4\eta-4}{4\eta}$, $\ds\frac{\eta+2}{4(n+1)}$, $\ds\frac{(\eta-2)(\eta+6)}{8\eta}$ \\[5pt]
\hline\hline
\end{tabular}\caption{The equilibrium points with associated eigenvalues}\label{table:Yeig} 
\end{table}

\begin{table}[ht]
\begin{tabular}{|c|c|c|} \hline\hline
Eq.\ point & Eigenvalue & Eigenvector $V = \left[V^{\Sigma},V^K,V^Y\right]$ \\
\hline
$L_1$  &  $-\Sigma$               &  $\left[1,0,0\right]$ \\
$L_1$  &  $-F\Sigma$              &  $\left[-(1-\Sigma^2),2\Sigma(1-\Sigma^2),-s\right]$ \\
$L_2$  &  $2y$                    &  $\left[2(1+2y),-12y,-s(1-Y)(1+2y)\right]$ \\
$P_1$  &  $\ds\frac2{(n+1)(\eta+2)}$  &
$\left[-2(\eta+2)(n\beta+2\eta-4),16(n+1)\beta,-(\eta+2)^3\left(\beta s^{-1}+2s+\eta+2\right)\right]$ \\
\hline\hline
\end{tabular}\caption{Eigenvectors of some important
  eigenvalues. Definitions: $\beta=\eta^2+4\eta-4$, $s=\eta/(n+1)$ }\label{table:eigvec}
\end{table}

\section{Computing the optical geometry using the loaf of bread system}
\label{sec:compute}

In this section we will use the previous analysis to study the 
\emph{optical geometry}\cite{aabgs:optical} of static spherically symmetric models. The optical geometry 
may be defined as the spatial part of the conformally rescaled original geometry, with 
a specific conformal factor. The symmetries allow us to visualize essentially all important
features as an embedded surface in a 3-dimensional Euclidean space. The geodesics of 
this surface then correspond to the null geodesics of the original spacetime which 
enable us to use our ordinary Euclidean intuition to investigate phenomena
such as closed lightlike orbits and their stability. 
We now derive the equations that describe the embedding.

We start from the Schwarzschild metric (\ref{SSSmetric})
and consider only the null geodesics, \emph{i.e.}~$\der s^2=0$ using the fact that 
conformal rescalings of the metric leave them invariant. We may thus multiply both sides 
by the non-zero conformal factor $e^{-2\nu}$. We also restrict attention, without loss of 
generality, to the equatorial plane $\theta=\pi/2$ thereby obtaining the new metric
\BE
0=\der \bar{s}^2=-\der t^2+e^{2(\lambda-\nu)}\der r^2+e^{-2\nu}r^2\der \phi^2
\EE
It is clear that the spatial geodesic equations will be independent of time, and that the 
timelike part will have a trivial solution, so we drop the time part of the metric and 
turn our interest to only the spatial part
\BE
\der l^2=e^{2(\lambda-\nu)}\der r^2+e^{-2\nu}r^2\der \phi^2 \ .  
\EE
This expression may be written in a standard form if we introduce two new variables given by
$\der r_*=e^{(\lambda-\nu)}\der r$ and $\tilde{r}=e^{-\nu}r$, yielding
\BE
\der l^2=\der r_*^2+\rtil^2\der \phi^2 
\label{geom}
\EE
so that radial distances in the optical geometry are measured by $\der r_*$, while distances 
along a circle of constant $r_*$ are measured by $\rtil\der \phi$. For this reason we refer
to $r_*$ as the \emph{optical radius} and $\rtil$ as the 
\emph{optical circumference variable}. Even though $r_*$ is a monotonous function of the 
Schwarzschild radius $r$, the relationship need in no way be simple, so it is not possible to 
directly read off $r$ from the optical geometry. For later reference we also point out that 
$\rtil$ has a simple connection to the centrifugal part of the effective potential, 
$V_l=l(l+1)/\rtil^2$.
We now wish to embed the optical geometry (\ref{geom}) in ${\mathbb{R}}^3$. To this end we write down the 
Euclidean metric in cylindrical coordinates.
\BE
\der s^2_E=\der z^2+\der \rhotil^2+\rhotil^2\der\phi^2
\label{Euclid3}
\EE
The embedded surface will, at least locally, 
be given by some functional relation $f(z,\rhotil)=0$, which, again locally, may be solved to 
give $z=h(\rhotil)$, where $h$ is some function. Thus $\der z=h_{,\rhotil}\der\rhotil$,
and the induced surface metric becomes
\BE
\der l^2=\left[1+\left(h_{,\rhotil}\right)^2\right]\der \rhotil^2+\rhotil^2\der\phi^2.
\label{Euclid2}
\EE
Comparing (\ref{Euclid2}) with (\ref{geom}) we find the relations
\BE
\der h^2=\der r_*^2-\der\rtil^2 \nonumber, \qquad \rhotil=\rtil(r_*), \
\label{embed}
\EE
which defines a differential equation for $h(\rho)$ given below.
\BE
\rtil=e^{-\nu}r, \qquad \frac{\der h}{\der r}=e^{-\nu}\sqrt{e^{2\lambda}-\left(1-\nu_{,r}r\right)^2}
\EE
In the interior of the star we may express this system in the state space variables of 
section 2. Thus extending the state space by the variables $\rtil$ and $h$ we are able to 
solve the embedding equations simultaneous with the dynamical system. The new structure 
equations are 
\BE
\dot{\rtil}=y\rtil(1-2\Sigma), \qquad \dot{h}=y\rtil\sqrt{K-(1-2\Sigma)^2}
\label{eq:evolution}
\EE It is obvious that the center of the star ($\rtil=0$) still
defines an invariant submanifold for this system, so we must again
linearize the equations to find initial conditions at a finite
distance from the center of the stellar object. The linearized
solution is in this case 
{\setlength\arraycolsep{2pt} \BE
  \rtil^2 & = & \epsilon\exp(2y_{\rm c}x) \nonumber \\
  h & = & \sqrt{2+y_{\rm c}}\epsilon\exp(2y_{\rm c}x), \EE} 
where we were forced
to linearize $\rtil^2$ rather than $\rtil$, since as $x\rarr-\infty$,
$\rtil^2$ is asymptotically a linear function of the original variables
$\Sigma$, $K$ and $Y$, whereas $\rtil$ itself is not. The solution
thus obtained may be scaled to any given mass of the star by matching
it to the appropriate exterior Schwarzschild solution at the surface.
In the exterior we have $e^{2\nu}=1-2\zeta$ and
$e^{2\lambda}=\left(1-2\zeta\right)^{-1}$, where $\zeta=M/r$, so the
embedding equations (\ref{embed}) take the form 
\BE \rtil=\frac{r}{\sqrt{1-2\zeta}},
\qquad \frac{\der h}{\der
  r}=\sqrt{\frac{\zeta\left(4-9\zeta\right)}{\left(1-2\zeta\right)^3}}. \EE  
From the fact that necks and
bulges correspond to the zeros of $\rtil_{,r}$ it is clear that
$\zeta=1/3$ correspond to a neck in the exterior geometry iff $R<3M$.
We may remark here that in the interior the necks and bulges
correspond to the zeros of $\Sigma-1/2$ as is easily seen from
(\ref{eq:evolution}).

\section{Optical geometries with an unlimited number of necks}
\begin{figure}
  \includegraphics[35mm,115mm][200mm,180mm]{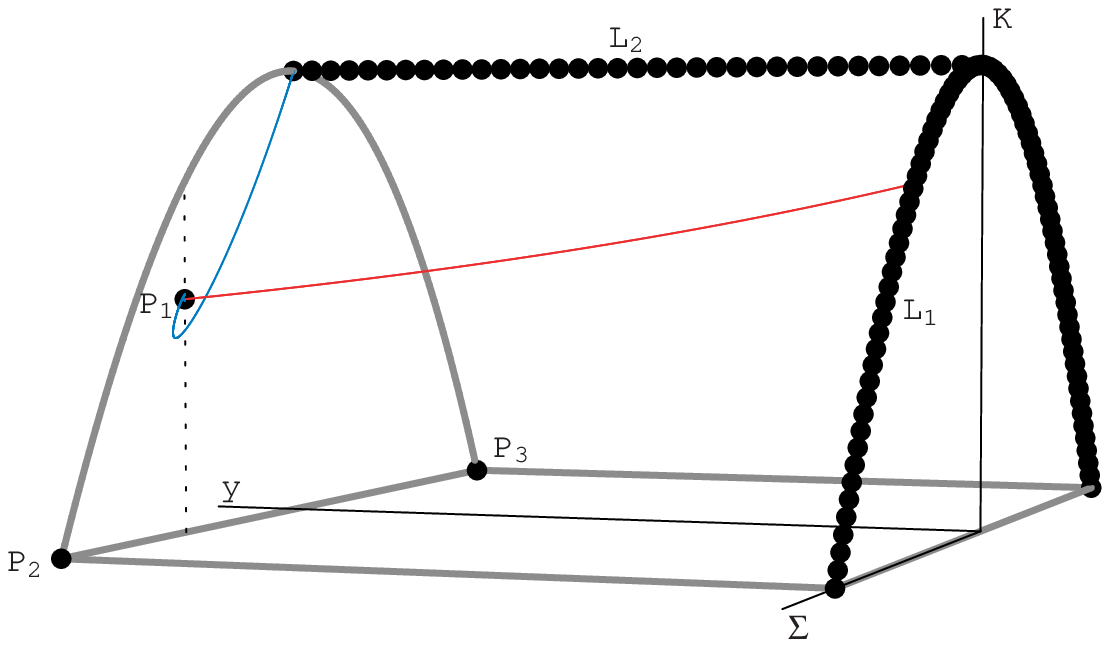}
  \caption{The state space of the loaf of bread system for the stiff equation of state $\rho=p+\rho_{\rm s}$. The equilibrium points, indicated as bold dots in the figure, are denoted as in table \ref{table:Yeig}. The figure also displays the combined limiting orbit described in section \ref{sec:regular}, corresponding to the limit of infinite central pressure within the family of regular center orbits. The RSI part of this combined orbit, which completely lies in the boundary $Y=1$, starts out from the $Y=1$ point on $L_2$ and spirals into $P_1$. However, from the figure it is not at all clear that the RSI orbit actually spirals into $P_1$, which is explained by the fact that the absolute value of the quotient $q$ between the imaginary and the negative real part of the relevant complex conjugate eigenvalues is not very large, the precise value being $q=\sqrt{3}$. This implies that the distance to $P_1$ decreases by a factor $e^{-2\pi/q} \approx 0.026$ for every turn around $P_1$. The skeleton part of the combined orbit, around which the surface of regular orbits is rolled, is for this particular equation of state confined to the plane $\Sigma=1/2$.}
\label{fig:stiff_loaf}
\end{figure}

Using the fact that an interior neck or bulge in the optical geometry
corresponds to the state space orbit crossing the surface
$\Sigma=1/2$, we now prove what was claimed in
\cite{rosquist:multneck}, namely the existence of causal models for
which the number of interior necks can be arbitrarily large. To see
this, we begin by reading off from table \ref{table:Yeig} that the
position of the Tolman equilibrium point $P_1$ on the $Y=1$ plane
depends on the equation of state parameter $\eta$. In particular,
since the $\Sigma$ coordinate of the point is $2/(\eta+2)$, $P_1$ can
be put on the $\Sigma=1/2$ surface by setting $\eta=2$, which puts it
at $\Sigma=K=1/2$ and gives a speed of sound that approaches that of
light in the limit of infinite pressure. The assertion now follows
almost immediately from our previous discussion of the surface of
regular orbits. Since $\eta=2$ exceeds the critical value
$\eta_*=-2+8/\sqrt{7}\approx 1.02$ above which the RSI orbit spirals into $P_1$,
an arbitrarily large number of $\Sigma=1/2$ crossings for a regular
and finite central density orbit can be obtained by choosing a
sufficiently large (but finite) central pressure, making the orbit
follow the combined RSI/skeleton limiting orbit sufficiently closely.
Although this proof does not depend on the particular form of the
pressure function $\psi$ in the equation of state, an arbitrarily
large number of necks may not always be obtained for models with
finite radii. However, since linear equations of state
$\rho=(\eta-1)p+\rho_{\rm s}$ ($\psi=\rho_{\rm s} = \rm const>0$) are
known to always give regular solutions with finite radii, we may
simply choose the stiff equation of state $\rho=p+\rho_{\rm s}$ to
avoid the possibility of ending up with models that are infinite in
extent at large central pressures. We also remark that for the stiff
equation of state, the eigenvector generating the skeleton solution
has a vanishing $\Sigma$ component, as can be read off from table
\ref{table:eigvec}. In fact, the skeleton solution is in this case an
exact solution which is entirely confined to the $\Sigma = 1/2$ plane,
namely the solution $\Sigma=1/2$, $K=\frac34(y+1)^{-1}$
\cite{tolman:sssfluids}.
This is likely to imply that the number of necks grows faster with the
central pressure than it does for more general forms of $\psi$, since a
nonvanishing $\Sigma$ component of course drives the skeleton orbit
away from $\Sigma=1/2$. The combined RSI/skeleton orbit for the
Zel'dovich equation of state is shown in fig.\ \ref{fig:stiff_loaf}.

\section{Some multiple neck optical geometries}
\label{sec:multneck}

Here we investigate the interior solution for different equations of
state. Our present purpose is to show that the interior optical
geometry will look quite different for the different cases. We showed
in an earlier paper \cite{krs:annalen} that the geometry may develop
additional ``necks'' and ``bulges'' corresponding to unstable and
stable closed lightlike orbits respectively.  As shown by
Chandrasekhar and Ferrari \cite{cf:osc} the axial modes of nonradial
metric perturbations of SSS perfect fluids are (to first order) not
coupled to a perturbation of the matter. Such metric perturbations may
therefore be interpreted as pure gravitational waves. By separation of
variables the equations governing these modes can be reduced to the
1-dimensional Schr\"odinger equation \BE -\frac{d^2}{dr_{\!*}^2}Z_{l}
+ V Z_{l} = \sigma^2 Z_{l} \EE where the total potential $V =
V_{l}+V_m$ is dominated by $V_{l} = l(l+1)\tilde{r}^{-2}$,
$l=2,3,\ldots$ which is simply the effective potential for photons
with angular momentum $L^2 = l(l+1)$.  This, together with the fact
that the correction $V_m$ to the total potential is small compared to
$V_l$ in most cases, shows that the optical geometry is closely
related to the trapped modes of gravitational radiation since the
maxima and minima of $V_l$ clearly correspond to the necks and bulges
of the optical geometry respectively.

We turn, now, to the analysis of the optical geometry of some different
models. In the case of a double layer uniform density model the equation 
of state is parametrized by the
constant densities $\rho_{+}$ and $\rho_{-}$ of the interior and
exterior zones respectively, as well as the transition pressure
$p_{\mathrm t}$ where the density is discontinuous. In the single layer uniform 
density model considered by Abramowicz \emph{et.\ al.}\ 
\cite{aabgs:optical} the only parameter is the density $\rho_0$, and the linear model 
is parametrized by the central and surface density $\rho_{\rm c}$ and
$\rho_{\rm s}$ respectively. 
Given these parameters, each stellar model is obtained by specifying the central
pressure $p_{\mathrm c}$.

Below we display and compare the optical geometry for these models. The parameter
values for the two-zone models are chosen to obtain three distinct examples of optical
geometries with pronounced double necks, corresponding to two deep
potential wells as shown in \cite{krs:annalen}. They were chosen to
have the same value of the tenuity $\alpha := R/M=2.2749$, as well as
the same mass. This condition is visualized by
letting the exterior geometry be identical in each triplet of
figures. The chosen parameters are shown in table \ref{tab:parameters},
where we have chosen to use cgs-units and assigned a mass of $1.5
M_\odot$ to each model. We see that the central density required
to produce the double neck structure is very large. We could have
chosen to have a more realistic central density, but then the total
mass of the corresponding stellar object would have been of the order of
several hundred solar masses.

\begin{figure}[ht]

 \begin{minipage}[t]{0.33\linewidth}
  \vspace{0pt}
  \centering 
  \includegraphics[scale=1]{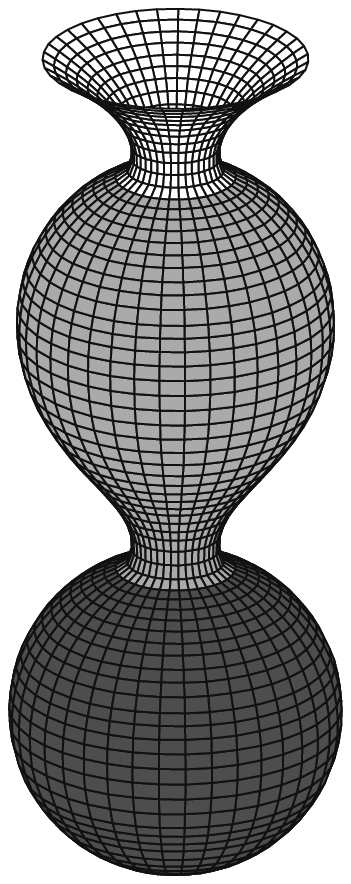}
 \end{minipage}  
\begin{minipage}[t]{0.33\linewidth}
  \vspace{0pt}
  \centering
  \includegraphics[scale=.442]{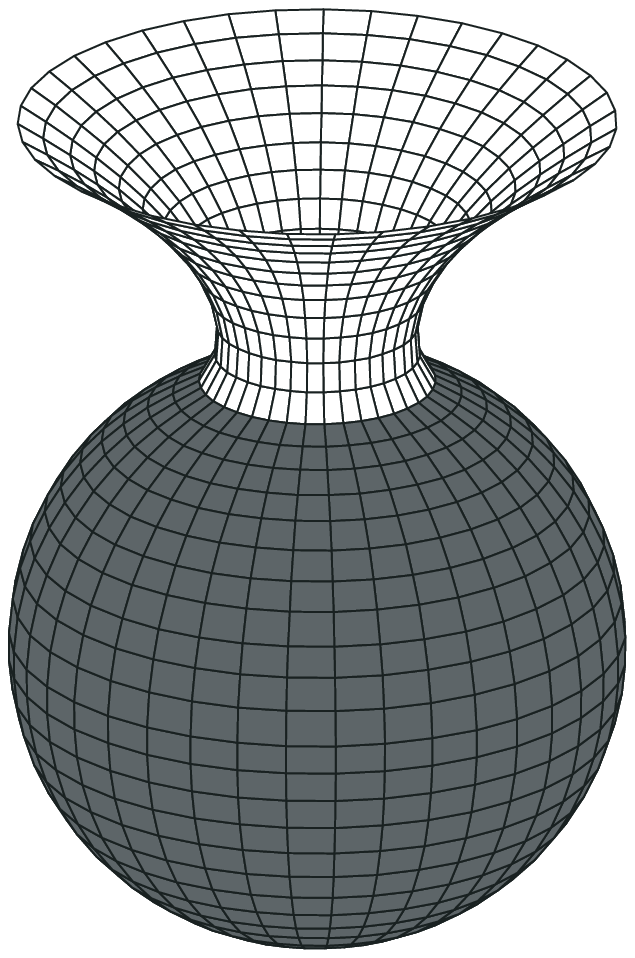}
 \end{minipage} 
 \begin{minipage}[t]{0.33\linewidth}
  \vspace{0pt}
  \centering
  \includegraphics[scale=.273]{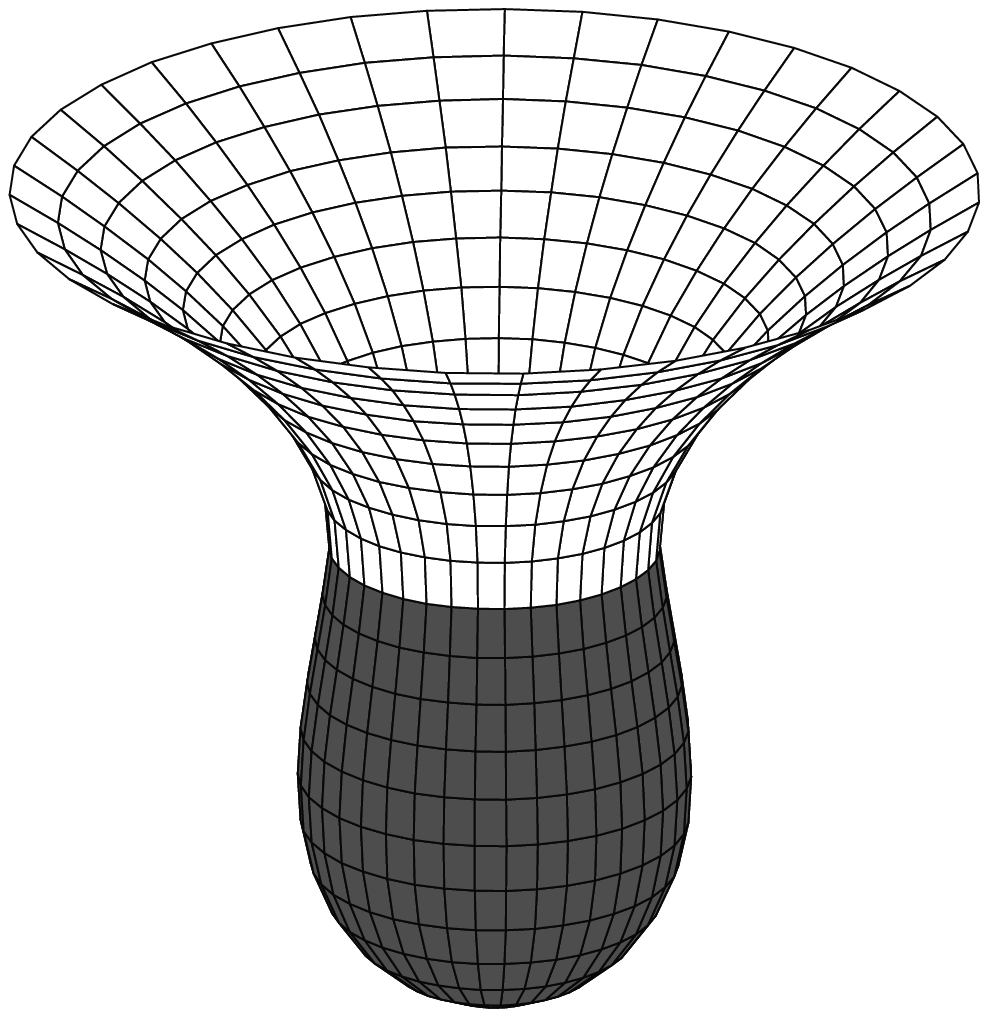}
 \end{minipage} 
 \caption{The optical geometry of the ``peanut'' model compared with
   the uniform and the linear models. All models have the same mass
   and hence identical exterior geometry.}
 \label{fig:comparison}
\end{figure}

In figure \ref{fig:comparison} we compare a two-zone model with a
uniform density model, and a model with a 
linear equation of state. The parameters of the two-zone model were
chosen so that the bulges should have approximately the same size
corresponding to two equally shaped potential wells. The density of
the uniform model were then chosen to give the same tenuity as this
model. The parameters of the linear model were chosen so that the
single neck should be reasonably visible. The tenuity of
this model is $\alpha=2.81$ which is close to the limit
$\alpha_{\mathrm min}=2.749$ for this type of equations of state. 
The lower part of the double layer model is of course, as in the
uniform case, a sphere. We thus expect that the modes of the inner
potential well, granted that the inner neck is thin, should resemble
those of the uniform model. 

\begin{figure}[ht]
 \begin{minipage}[t]{0.33\linewidth}
  \vspace{0pt}
  \centering
  \includegraphics[scale=.633]{peanut.eps}
 \end{minipage}  
\begin{minipage}[t]{0.33\linewidth}
  \vspace{0pt}
  \centering
  \includegraphics[scale=.444]{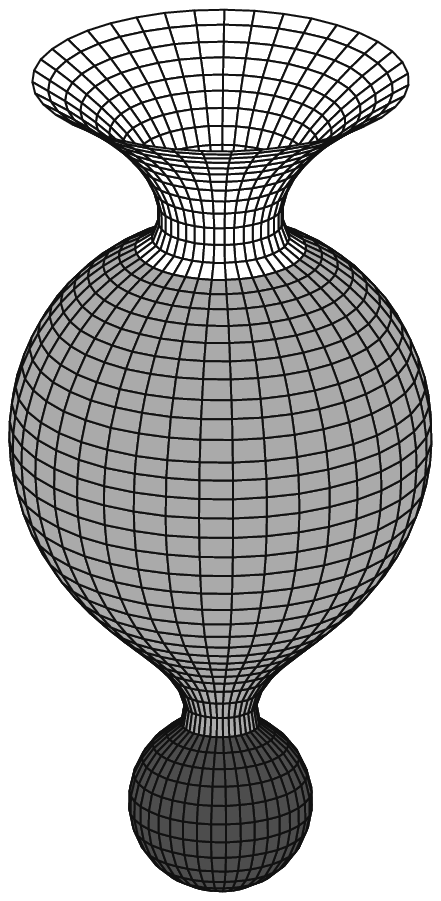}
 \end{minipage} 
 \begin{minipage}[t]{0.33\linewidth}
  \vspace{0pt}
  \centering
  \includegraphics[scale=.9]{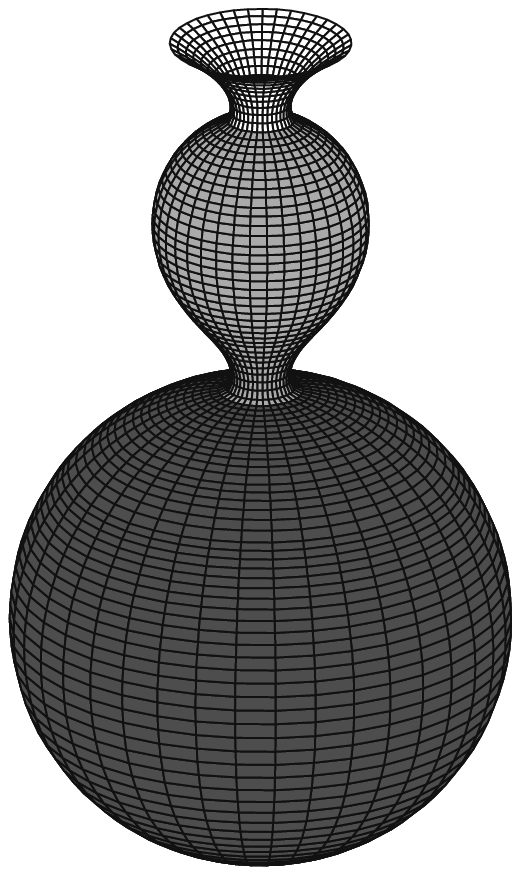}
 \end{minipage} 
 \caption{The optical geometry of the ``peanut'' model compared with
   the ``bigup'' and ``bigdown'' models. All models have the same mass
   and hence identical exterior geometry.}
 \label{fig:twozone}
\end{figure}

In figure \ref{fig:twozone} we look at three different two-zone
models. The first is the one already shown in figure
\ref{fig:comparison}. The two other models are chosen to display the
possibility to have either a large upper bulge or a large lower one. 

It would be interesting to know if the two-zone models are stable
against small radial perturbations, but this question turns out to 
be nontrivial. Of course, a star with constant density is always
stable, however since this configuration is very unphysical (the speed 
of sound diverges) one usually views the equation of state $\rho$ =
constant as an approximation to the more physical
$\rho=\rho_0+\rho_1(p)$, where $\rho_0$ is a constant and $\rho_1(p)$
is a function of the pressure satisfying $\rho_1(p)\ll \rho_0$ for all 
$p<p_{\rm c}$. One may then impose reasonable conditions on this equation of 
state, for instance causality ($v_{\rm{sound}}\le 1$), and the dominant energy
condition ($\rho\le p$). It seems however that you cannot find any
reasonable functions $\rho_1(p)$ such that these conditions hold and
the star is ultrarelativistic. We may note that none of our two-zone
models respects the dominant energy condition. Ignoring these
physical difficulties and viewing the models as toys, we may proceed
with the stability analysis.
Firstly we may remark that you cannot use the usual
analysis of the mass to radius curve to establish stability properties 
for any model with constant density since the adiabatic index is not
defined \cite{thorne:struct} (it  is determined by the small function
$\rho_1(p)$ which is arbitrary here). We are therefore forced to try a
dynamical approach, but here the analysis is complicated by the
discontinuity in the perturbation equation so that the differential
equation is no longer of the standard Sturm-Liouville type. Therefore, 
the standard theorems of Sturm-Liouville theory cannot be directly
applied. However, since the discontinuous density function can be
arbitrarily well approximated by
a continuous function one might expect that nothing too extraordinary
happens. Preliminary results using a variational principle indicate
that this is indeed the case. The stability of more realistic
ultracompact (in the sense $R\le 3M$) stellar configurations with a phase
transition has previously been studied by Iyer, Vishvehvara and
Dhurandhar \cite{ivd:ultracompact}. They conclude that there exist
stable causal ultracompact models, however not with any of the
high density equations of state proposed in the litterature at that
time. 

\begin{table}[ht]
\begin{tabular}{|c|c|c|c|c|c|} \hline\hline
Two-zone models &$\rho_+ (g/cm^3)$  & $\rho_- (g/cm^3)$ & $p_{\rm t} (g/cm s^2)$ 
& $p_{\rm c} (g/cm s^2)$ & $r_{\rm t}/R$ \\
\hline
Peanut & $2.74\times 10^{19}$ & $5.48\times 10^{15}$ & $3.52\times 10^{38}$ 
& $5.67\times 10^{41}$ & 0.014 
\\
\hline
Bigup & $5.81\times 10^{19}$ &  $5.51\times 10^{15}$ & $2.90\times 10^{38}$ 
& $5.22\times 10^{41}$ & 0.010 
\\
\hline
Bigdown & $2.80\times 10^{19}$ & $5.48\times 10^{15}$ & $3.63\times 10^{38}$ 
& $2.90\times 10^{42}$ & 0.014 
\\
\hline\hline
\end{tabular}
\\[0.3cm]
\begin{tabular}{|c|c|c|c|}\hline\hline
Other models & $\rho_{\rm c} (g/cm^3)$  & $\rho_{\rm s} (g/cm^3)$ &
$p_{\rm c} (g/cm s^2)$ \\
\hline
Uniform density & $5.56\times 10^{15}$ & & $7.59\times 10^{37}$ \\
\hline
Linear & $9.00\times 10^{16}$ & $5.56\times 10^{15}$ & $7.59\times 10^{37}$ \\
\hline\hline
\end{tabular} 
\caption{Parameter values for the models considered. Here we have
  chosen the mass of the object to be $1.5M_\odot$ which means that the 
  radius will be about $6.2$ km for the linear model, and $5.0$ km for 
  the others. We see that the term ultracompact is indeed relevant
  since these objects are denser than normal neutron stars are
  believed to be. We see that the two-zone models consist of 
  very dense, small cores ($r_{\rm t}$ refers to the radius where the phase
  transition takes place), and envelopes of more realistic
  densities.} 
\label{tab:parameters}
\end{table}

\section{Concluding remarks}
\label{sec:conclude}

The two-zone models discussed in this paper could be considered as toy 
models for stars having a first order phase transition in their
interiors.  Our results indicate that a phase transition is an important
mechanism for producing a pronounced neck in the optical geometry of
a stellar model. In this picture, the outer neck at $r=3M$ (if it
exists) is associated with the vacuum/matter
phase transition at the surface of the star. However, this is not the
only mechanism for producing necks. Another example is provided by the 
stiff linear equation of state. In that case there is an increasingly
long sequence of shallow necks which is associated with the high density
spiral equilibrium point $P_1$\cite{rosquist:multneck}.

The existence of multiple neck optical geometries is associated
with gravitational perturbation potentials with multiple wells \cite{krs:annalen}.  It would
therefore be of interest to calculate the corresponding quasi-normal modes
along the lines of calculations done for single well potentials in
\cite{cf:gwresonance} (uniform density models) and in \cite{ak:polytropes}
(polytropic models).

\section*{Acknowledgement}
We would like to thank Claes Uggla for helpful remarks on the manuscript.


\bibliographystyle{prsty}
\bibliography{max}

\begin{thebibliography}{10}

\bibitem{cf:gwresonance}
S. Chandrasekhar and V. Ferrari, \prsla {\bf 434},  449  (1991).

\bibitem{aabgs:optical}
M.~A. Abramowicz {\it et~al.}, \cqg {\bf 14},  L189  (1997).

\bibitem{rosquist:trapped}
K. Rosquist, \prd {\bf 59},  044022  (1999), (\eprint{gr-qc/9809033}).

\bibitem{rosquist:multneck}
M. Karlovini, K. Rosquist, and L. Samuelsson, Ultracompact stars with multiple
  necks, 2000, \eprint{gr-qc/009073}.

\bibitem{krs:annalen}
M. Karlovini, K. Rosquist, and L. Samuelsson, Annalen der Physik {\bf 9},  149
  (2000).

\bibitem{nu:grstars_linear}
U.~S. Nilsson and C. Uggla, General relativistic static stars: Linear Equations
  of State, 2000, \mbox{\eprint{gr-qc/0002021}}.

\bibitem{nu:grstars_polytropic}
U.~S. Nilsson and C. Uggla, General relativistic static stars: Polytropic
  Equations of State, 2000, \mbox{\eprint{gr-qc/0002022}}.

\bibitem{lindblom:phase}
L. Lindblom, \prd {\bf 58},  024008  (1998).

\bibitem{lm:indexlimit}
L. Lindblom and A.~K.~M. {Masood-ul-{A}lam},  in {\em Directions in General
  Relativity II: Papers in Honor of {D}ieter {B}rill}, edited by B.~L. Hu and
  T.~A. Jacobson (Cambridge University Press, Cambridge, 1993), p.\ 172.

\bibitem{ms:msmass}
C.~W. Misner and D.~H. Sharp, \pr {\bf 136},  571  (1964).

\bibitem{htww:gravcollapse}
B.~K. Harrison, K.~S. Thorne, M. Wakano, and J.~A. Wheeler, {\em Gravitation
  theory and gravitational collapse} (University of Chicago Press, Chicago,
  USA, 1965).

\bibitem{mtw:gravitation}
C.~W. Misner, K.~S. Thorne, and J.~A. Wheeler, {\em Gravitation} (Freeman, San
  Fransisco, USA, 1973).

\bibitem{tolman:sssfluids}
R.~C. Tolman, \pr {\bf 55},  364  (1939).

\bibitem{cf:osc}
S. Chandrasekhar and V. Ferrari, \prsla {\bf 432},  247  (1991).

\bibitem{thorne:struct}
K.~S. Thorne,  in {\em Proccedings of the international school of physics
  ``Enrico Fermi''}, edited by L. Gratton (Academic Press, New York and London,
  1966), p.\ 166.

\bibitem{ivd:ultracompact}
B.~R. Iyer, C.~V. Vishveshwara, and S.~V. Dhurandar, \cqg {\bf 2},  219
  (1985).

\bibitem{ak:polytropes}
N. Andersson and K.~D. Kokkotas, \mnras {\bf 297},  493  (1998).

\end{thebibliography}

\end{document}